\documentclass[10pt,5p]{elsarticle}
\usepackage{color}
\usepackage{amsmath,amsthm}
\usepackage{graphicx}
\usepackage{dcolumn}
\usepackage{bm}
\usepackage{color}
\usepackage{epsfig}
\usepackage{multirow}

\definecolor{red}{rgb}{0.71,0.14,0.07}

\def\expG{  { e^{- \Xi c( {\bf r}  ) /2  }  } }

\def\Tr{{{\rm Tr}\; }}

\def\r{{{\bf r} }} 
\newcommand{\mb}[1]{\mathbf{#1}}

\journal{Journal of Computational Physics}

\begin{document}

\begin{frontmatter}



\title{Solving Fluctuation-Enhanced Poisson-Boltzmann Equations}



\author{Zhenli Xu} \ead{xuzl@sjtu.edu.cn}
\address{Department of
  Mathematics, Institute of Natural Sciences, and MoE Key Lab of
  Scientific and Engineering Computing, Shanghai Jiao Tong University, 800 Dongchuan Rd.,
  Shanghai 200240, China }
\author{A. C. Maggs}\ead{anthony.maggs@espci.fr}
\address{Laboratoire de
  Physico-Chime Th\'eorique, ESPCI, CNRS Gulliver, 10 Rue Vauquelin,
  75005,Paris,France}

\begin{abstract}
  Electrostatic correlations and fluctuations in ionic systems can be
  described within an extended Poisson-Boltzmann theory using a
  Gaussian variational form.  The resulting equations are challenging to solve
  because they require the solution of a non-linear partial
  differential equation for the pair correlation function.  This has
  limited existing studies to simple approximations or to
  one-dimensional geometries. In this paper we show that the numerical
  solution of the equations is greatly simplified by the use of
  selective inversion of a finite difference operator which occurs in
  the theory. This selective inversion preserves the sparse structure
  of the problem and leads to substantial savings in computer
  effort. In one and two dimensions further simplifications are made
  by using a mixture of selective inversion and Fourier techniques.
\end{abstract}


\end{frontmatter}







\section{Introduction}

Electrostatic phenomena are important in many fields of engineering,
physical and biological sciences. Examples where their study is
crucial to the understanding of the large scale properties of
materials include colloidal suspensions, electrochemical energy
devices, DNA and membrane function, and nano-particle interactions
\cite{BKNNSS:PP:05,FPPR:RMP:10,Naji:13}. Classical Poisson-Boltzmann theory
\cite{Gouy:JP:10,Chapman:PM:13} is a widespread approach for
understanding phenomena in such systems, but it is a mean-field
approach which determines the ion concentrations by the average
electrostatic potential; it ignores many-body effects and
fluctuations. One simple example of the weakness of the standard
Poisson-Boltzmann is given by the example of an air-water interface
where the equations eliminate self-energy effects and image charges
\cite{Wang:PRE:10,Kanduc:EPJE:07}. The density profile predicted by the equations is
then a poor approximation close to the surface.

One measure of many-body effect in charged systems is the coupling
parameter $\Xi$, \cite{netz:2001}, an a-dimensional parameter
calculated from surface charge density, $\sigma e$, counter-ion valency
$q$, dielectric permittivity $\epsilon $ and inverse temperature
$\beta$. In SI units, $$\Xi= q^3 |\sigma| e^4 \beta^2/8 \pi \epsilon^2, $$ with $e$
the electronic charge.  In the weak-coupling regime, $\Xi\rightarrow
0$; the mean-field, Poisson-Boltzmann theory works rather well. In the
strong-coupling limit, $\Xi\rightarrow \infty$, other theoretical
approaches have been developed starting from a single ion picture
\cite{loth:2009,BKNNSS:PP:05,samaj:2011}.  This strong-coupling theory
successfully captures phenomena such as counterion condensation, and
like-charge attraction. The perturbative expansion in either small
$\Xi$ or large $\Xi$ can not be applied however for the intermediate
regime $\Xi\sim 1$, which is however often found experimentally.

By the use of a Gaussian variation {\it ansatz}, Netz and Orland
\cite{NO:EPJE:00,NO:EPJE:03} derived a system of self-consistent
partial differential equations which are believed useful for a larger
range of $\Xi$. Historically, the equations were also derived by
Avdeev and Martynov using the Debye closure of the BBGKY hierarchical
chain \cite{AM:CJ:86}.  Based on the self-consistent equations, there
has been recent interest of numerical methods and analysis with
one-loop expansion or asymptotic approximation; See Buyukdagli {\it et
  al.}  \cite{BMP:PRE:10,BAA:JCP:12}, and references therein; these
studies show excellent agreement with Monte Carlo simulations.  The
variational field-theoretical approach has been discussed by many
other authors: \cite{Yaroshchuk:ACIS:00,HL:SM:08,LLPS:PRE:02}.
Clearly the equations contain much useful physics and it would be
useful to have a tool to study them in more general geometries.

The main result of the present paper is the presentation of numerical
methods that enable one to {\it solve}\/ the self-consistent equations
in general geometries. To do this we firstly discretise the
variational equations. When we do so we see that the main technical
difficulty that has to be overcome is the inversion of a large system
of linear equations. However, closer examination shows that we only
require diagonal elements of the inverse matrix, this diagonal inverse
is related to a local electrostatic energy of an ion, similar to a
spatially varying Born energy \cite{Wang:PRE:10}. The matrix equations
in the self-consistent formulation are {\it sparse}\/ in structure --
that is, they are matrices for which almost all entries are zero.  We
thus use techniques developed to {\it selectively invert}
\cite{LYMLYE:ACMMS:11,linlin} the matrices which occur in our problem. This
selective inversion requires less memory storage and also less
computer operations to perform than the usual algorithms which
calculate the full, dense solution of the linear system. This selected
inversion is then used as part of an iterative solver to find the
self-consistent solution of the full extended Poisson-Boltzmann
equations.

The small workstation which we used for our implementation limits us
to modest system sizes in arbitrary three-dimensional
geometries. However, if we study physical systems which are invariant
in one, or two dimensions we can do much better. We give as an example
of the interaction between two charged cylinders (2+1 dimensions), or
the profile of density near a planar wall (1+2 dimensions). In these
cases we can combine the method of selective inversion with Fourier
analysis in the invariant directions. Then rather fine discretisations
are possible even with modest computational resources.  We find that
we are able to then generate numerical solutions of the equations for
$0<\Xi<4.7$, which includes rich physics, but does not reach the
regime of like-charge attraction.

To be fully useful in three dimensions, more efficient, multiprocessor,
codes will be needed. However, several groups are
working on scalable selected inversion \cite{MUMPS:2,linlin}. We can
expect rather rapid progress on the application to three dimensions.
We note that the utility of sparse matrix methods for free energy
calculations has already been noted in the calculation of fluctuation
based interaction such as van der Waals interactions
\cite{pasquali:2008,pasquali:2009}.

\section{Self-consistent equations for symmetric electrolytes}

We consider a symmetric electrolyte composed of two ion species of
valence $\pm q$, in a dielectric medium with fixed charge density
$\rho_f(\mathbf{r})$. We assume the ions are point charges, and the
ion accessible region $\Omega$ has uniform dielectric permittivity;
excluded volume effects \cite{BAO:PRL:97} are ignored in our present
paper. We will only allow ions to explore a region with constant
dielectric background in order to simplify the treatment of the self
energy. We will allow, however, non-fluid regions to have varying
dielectric properties in order to describe the electrostatics of
membranes or electrodes.

The following two length scales are important for electrolytes. The
Bjerrum length, $\ell_B=e^2/(4\pi\varepsilon_0\varepsilon_wk_BT)$,
gives the distance at which two unit point charges have an interaction
$k_BT$. Here, $\varepsilon_0$ is the vacuum dielectric permittivity,
$\varepsilon_w$ is the relative permittivity of water, and the product
$k_BT$ is the thermal energy. The Gouy-Chapman length,
$\ell_{GC}=1/(2\pi\ell_B q |\sigma_s|)$, describes the interaction
between a charge and a surface charge of density $\sigma_s e$, i.e., at
this length the interaction energy is comparable $k_BT$. Here we
suppose that $\sigma_s e$   is the
average surface charge density of the charge source
$\rho_f(\mathbf{r})$. We note that the coupling parameter can be
expressed via a ratio of the lengths, $\Xi=q^2 \ell_B/\ell_{GC}$.

We use dimensionless units by using the natural length $\ell_{GC}$, and
scaling the surface charge density to $\sigma_s
\ell_{GC}^2$. In the variational field theory, the grand-canonical
partition function is represented by the functional integral
$Z=\frac{1}{Z_G}\int \mathcal{D}\phi\exp\{-H[\phi]/\Xi \}$, where
$Z_G$ is a normalization factor, $i\phi$ is the fluctuating
electrostatic potential and the Hamiltonian functional is
\cite{NO:EPJE:03},
\[
H[\phi]=\int
\frac{d\mathbf{r}}{2\pi}\left[\frac{(\nabla\phi)^2}{4}+i\phi\rho_f-
  \frac{ \Lambda}{2} e^{ \Xi G_0(\mathbf{r},\mathbf{r})/2} \cos\phi
\right].
\]
Here, the ions contribute only in the accessible region $\Omega$,
$\Lambda=2\lambda/(\pi\sigma_s^2\ell_B)$ is the rescaled fugacity, and
$\lambda$ is a constant related to the chemical potential of symmetric
ions. We will also introduce
$G_0(\mathbf{r},\mathbf{r}')=1/|\mathbf{r}-\mathbf{r}'|$, the bare
Coulomb potential, in a uniform dielectric background. We have already
implicitly regularised the theory, since $G({\bf r},{\bf r}) $ is
divergent in the continuum limit.  The term in $\Xi
G_0(\mathbf{r},\mathbf{r})/2$ shifts the zero of the chemical
potential of the ions.

The variational method uses a general Gaussian Hamiltonian and
optimises over the mean and covariance. Explicitly we write,
\[
H_0[\phi]=\frac{1}{2}\int\int d\mathbf{r}d\mathbf{r}'
[\phi(\mathbf{r})+i\Phi(\mathbf{r})]{(\Xi G)}^{-1}(\mathbf{r},\mathbf{r}')[\phi(\mathbf{r}')+i\Phi(\mathbf{r})'],
\]
with $\Phi$ the mean potential, and the Green's function ${G}$ is the
covariance. The estimated Gibbs free energy is given by
$$\mathcal{F}_\mathrm{Gibbs}=\mathcal{F}_0+\langle H-H_0 \rangle_0/\Xi,$$
where $\mathcal{F}_0=-\frac{1}{2}\mathrm{Tr} \log G$ and
$\langle\cdot\rangle_0$ is the average with respect to $H_0$. One
finds,
\begin{align}
  &\mathcal{F}_\mathrm{Gibbs}=-\frac{1}{2}\mathrm{Tr} \log (\Xi G)-\int d\mathbf{r} \left[\frac{(\nabla\Phi)^2}{8\pi\Xi}-\frac{\rho_f \Phi}{2\pi\Xi}\right] &\nonumber \\
  &~~~~~~~~ +\int\int d\mathbf{r}d\mathbf{r}' \delta(\mathbf{r},\mathbf{r}')\frac{\nabla_\mathbf{r}\nabla_{\mathbf{r}'}G(\mathbf{r},\mathbf{r}')}{8\pi} & \nonumber\\
  &~~~~~~~~ -\frac{\Lambda}{4\pi\Xi}\int d\mathbf{r} e^{- \Xi
    [G(\mathbf{r},\mathbf{r})-
    G_0(\mathbf{r},\mathbf{r})]/2}\cosh\Phi. & \label{gibbs}
\end{align}
In this expression we understand the interest in introducing the
function $G_0$ so that the continuum limit is less singular -- we
expect that the combination $(G({\bf r}, {\bf r}) -G_0({\bf r}, {\bf
  r})) $ remains finite.

From the first variation of the free energy functional with respect to
the mean potential $\Phi$ and the Green's function
$G(\mathbf{r},\mathbf{r}')$, we obtain the following dimensionless
equations (see Appendix~\ref{appena} for the derivation),
\begin{subequations}\label{multieqn}
  \begin{align}
    &\nabla^2 \Phi - \Lambda e^{-\Xi c(\mathbf{r})/2 } \sinh \Phi =-2\rho_f(\mathbf{r}), \label{multieqn-a}\\
    &\left[\nabla^2  - \Lambda  e^{-\Xi c(\mathbf{r})/2 } \cosh \Phi \right]G(\mathbf{r},\mathbf{r}')=-4\pi \delta(\mathbf{r}-\mathbf{r}'),   \label{multieqn-b}\\
    &c(\mathbf{r})=\lim\limits_{\mathbf{r}\rightarrow
      \mathbf{r}'}\left[G(\mathbf{r},\mathbf{r}')-G_0(\mathbf{r},\mathbf{r}')\right], \label{multieqn-c}
  \end{align}
\end{subequations}
where the space-dependent function $\Xi c(\mathbf{r})/2$ describes the
effective self energy of a mobile ion. Since we will solve the
variational equations in a discretized form we derive them using the
language of matrices and lattice Green functions in Appendix~\ref{appena}.

Eq.~\eqref{multieqn-a} is the modified Poisson-Boltzmann equation
where the second term in the left hand side is the density of mobile
charge and the right hand side is the density of the fixed
charge. These equations are beyond the traditional Poisson-Boltzmann
theory.  The correlation function $c(\mathbf{r})$ represents the
electrostatic energy of a mobile ion interacting with surrounding
ions, which is determined by Eqs.~\eqref{multieqn-b} and \eqref{multieqn-c} self-consistently.
Eq.~\eqref{multieqn-b} is a semilinear differential equation in the
form of a modified Debye-H\"uckel equation, where the inverse Debye
length, found from $\kappa^2={\Lambda e^{-\Xi c(\mathbf{r})/2 }\cosh
  \Phi}$, depends nonlinearly on $c$ and $\Phi$.

The properties of the system are characterized by two parameters: the
coupling parameter $\Xi=q^2\ell_B/\ell_{GC}$ and the rescaled fugacity
$\Lambda=8\pi\lambda\ell_{GC}^3\Xi$. We can see that the equations can
be decoupled in the small $\Xi$ limit, when they reduce to the
classical Poisson-Boltzmann equation. The fugacity is actually a
measure of ratio between the Debye screening length and the
Gouy-Chapman length, and nonlinear effects come out for $\Lambda\ll 1$
where the Debye-H\"uckel theory does not work and the generalised
equation Eq. \eqref{multieqn-b} should be adopted.

\section{The two-level iterative algorithm}

In this section, we discuss the self-consistent iterative scheme,
and the algorithm for the modified Poisson-Boltzmann equation.

In the region outside $\Omega$ where we suppose the dielectric
permittivity is $\varepsilon_d$, the equations for the potential and
the Green's function are decoupled,
\begin{equation}\left\{ \begin{array}{ll}
      \eta_\varepsilon\nabla^2 \Phi =-2\rho_f(\mathbf{r}), \\
      \eta_\varepsilon\nabla^2 G(\mathbf{r},\mathbf{r}')=-4\pi \delta(\mathbf{r}-\mathbf{r}'),
    \end{array}\right. \end{equation}
where $\eta_\varepsilon = \varepsilon_d/\varepsilon_w$ is the relative
permittivity.  We define the indicator function of the ion accessible
region,
\begin{equation}
  \chi(\mathbf{r}) = \left\{ \begin{array}{ll}
      1, ~ ~ ~\hbox{if}~~\mathbf{r}\in\Omega, \\
      0, ~ ~ ~\hbox{otherwise},
    \end{array}\right.\end{equation}
and define the relative dielectric function of the space by using the
bulk water permittivity $\varepsilon_w$, i.e.,
$$\eta(\mathbf{r})=\eta_\varepsilon + \left(1-\eta_\varepsilon\right)\chi. $$
Then the governing equation for the whole space can be written as,
\begin{equation}\left\{ \begin{array}{ll}
      \nabla\cdot\eta(\mathbf{r})\nabla \Phi - \chi \Lambda e^{-\Xi c(\mathbf{r})/2 } \sinh \Phi =-2\rho_f(\mathbf{r}), \\
      \left[\nabla\cdot\eta(\mathbf{r})\nabla  - \chi \Lambda  e^{-\Xi c(\mathbf{r})/2 } \cosh \Phi \right]G=-4\pi \delta(\mathbf{r}-\mathbf{r}'), \\
      c(\mathbf{r})= \lim\limits_{\mathbf{r}\rightarrow \mathbf{r}'}\left[G(\mathbf{r},\mathbf{r}')-G_0(\mathbf{r},\mathbf{r}')/\eta(\mathbf{r})\right],
    \end{array}\right. \label{dimensionless}\end{equation}
where the continuity conditions across the interface $\partial\Omega$
are implied by the operator $\nabla\cdot\eta(\mathbf{r})\nabla$.


We develop a self-consistent iterative scheme for the solution of the
partial differential equations \eqref{dimensionless}. The iterative
scheme is composed of two alternating steps: For given $c({\bf r})$,
solve the modified Poisson-Boltzmann equation (the first equation) for $\Phi$
subject to given boundary conditions; and for given $c({\bf
  r})$ and $\Phi$, solve the modified Debye-H\"uckel equation (the
second equation) for $G$ and then a new $c({\bf r})$. For convenience,
we call them the PB and DH steps, respectively. These two steps are
iteratively performed until the convergence criteria of the solution
is reached. Mathematically, the iterative scheme is expressed by,
\begin{equation}\left\{ \begin{array}{ll}
      \nabla\cdot\eta(\mathbf{r})\nabla \Phi^{(k+1)} - \Lambda e^{-\frac{\Xi c^{(k)}}{2} } \sinh \Phi^{(k+1)}=-2\rho_f(\mathbf{r}),  \\
      \left[\nabla\cdot\eta(\mathbf{r})\nabla  - \Lambda  e^{-\frac{\Xi c^{(k)}}{2} } \cosh \Phi^{(k+1)} \right]G^{(k+1)}=-4\pi \delta(\mathbf{r}-\mathbf{r}'),\\
      c^{(k+1)}(\mathbf{r})=\lim\limits_{\mathbf{r}\rightarrow
        \mathbf{r}'}\left[G^{(k+1)}(\mathbf{r},\mathbf{r}')-G_0(\mathbf{r},\mathbf{r}')/\eta(\mathbf{r})\right],
    \end{array}\right.\end{equation}
for $k=0,1,\cdots, K$, where the superscript $(k)$ represents the
$k$th iteration step. The stop criteria
is $$\max|\Phi^{(K)}-\Phi^{(K-1)}|<\delta,$$ where $\delta$ is a small
pre-set value.

We see, at the $k$th step, the modified Poisson-Boltzmann equation is
still a nonlinear equation, which will be solved by iterative scheme,
too, therefore the numerical algorithm for the self-consistent
equations is a two-level iterative scheme. The modified Debye-H\"uckel
equation becomes an equation which can be solved by an explicit
algorithm, which will be discussed in the next section.

We present the sub-level iterative algorithm, used to solve the modified
Poisson-Boltzmann equation.  The method in the PB step has been widely studied in
literature \cite{LZHM:CCP:08,LLPA:MBMB:13}. We use an iterative algorithm based on
a three-point finite-difference discretisation for second-order differentiations of variable coefficients
\cite{IBR:CPC:98}, which remains the continuity of electric displacement across the
interface. We write the equation in the form,
\begin{equation}
  \nabla\cdot\eta(\mathbf{x})\nabla \Phi  - f(\mathbf{x}) \sinh \Phi =-2\rho_f(\mathbf{x}),
\end{equation}
with $f>0$ being space-dependent function, and $\mathbf{x}$ is a $d$ dimensional variable.

Let $h$ be the mesh size. The $d$-dimensional mesh is composed of $n^d$ lattice
sites where $n$ is the number of grid points in each direction.  We
define by $\Psi$ the vectorization of the potential values at lattice
sites and define by $H_1,\cdots, H_d$ the vectorization of the
dielectric permittivity at half integer grids at corresponding
dimension and integer grids at other dimensions, with periodic
boundary conditions. For example, in two dimensions,
$$\Psi=\{\Phi_{ij}, i, j=1,\cdots, N\},$$
and
$$\begin{array}{ll}
H_1=\{\eta_{i+1/2, j}, i,
j=1,\cdots, n\},\\
 H_2=\{\eta_{i, j+1/2}, i, j=1,\cdots, n\}.
 \end{array}$$ We
suppose these vectors are column vectors.

Let $\mathbf{D}$ and $\mathbf{D}^T$ be the difference matrices of
operators $\nabla\cdot$ and $-\nabla$, where
$\mathbf{D}=[\mathbf{D}_1,\cdots, \mathbf{D}_d]$ is of $n^d\times
dn^d$ dimensions. Then we can discretize the modified
Poisson-Boltzmann equation by,
\begin{equation}
  \left[-\mathbf{D} \mathrm{H} \mathbf{D}^T -\mathrm{diag}\{ \Gamma^{(l) } \}  \right]\Psi^{(l+1)} = R^{(l)},
\end{equation}
where $\mathrm{H}=\mathrm{diag}\{H_1^T,\cdots, H_d^T\}$ and
$\mathrm{diag}\{\cdot\}$ is the diagonal matrix with the vectorial
argument as the main diagonals, $\Gamma$ is the vectorial
representation of the relaxation function $$ \gamma(\mathbf{x}) =
f(\mathbf{x}) \left|\sinh\Phi \right| /\left( |\Phi
  |+\epsilon\right)\geq0,$$ with $\epsilon=10^{-8}$, and $R$ is the
vector for the function on the right hand side,
$$r(\mathbf{x})=f \sinh \Phi - \gamma \Phi-2\rho_f.$$

In real problems, the fixed charge $\rho_f$ are often point or surface
charge, and should be distributed into the lattice sites. In our
calculations, the surface charge (for instance on a curves surface) is
first discretised into point charges, say of fractional valencies,
$q_m$, $m=1,\cdots, M$. Each point charge is distributed into the
nearest grid sites by a linear weighting function \cite{IBR:CPC:98},
\begin{equation}
  \Delta \rho_{i_1\cdots i_d}=q_m \prod_{l=1}^d \left(1-\frac{|x_{m,l}-x_{i_l}|}{h}\right),
\end{equation}
where $x_{m,l}$ is the $l$th coordinate of the $m$th charge, and
$x_{i_l}$ is the $l$th coordinate of one of the nearest lattice sites.
Then $\rho_{i_1\cdots i_d}$ is the superposition of all fractional
contributions of point source charges.

The discretisation yields symmetric and negatively definite
coefficient matrix, and thus the equations can be efficiently solved
by standard direct solvers. We use sparse matrix division in Matlab in our
implementation.

\section{The generalized Debye-H\"uckel equation}

The greatest difficulty in solving the self-consistent equations arises from the modified DH equation, which
can be reformulated during the self-consistent iteration as,
\begin{equation}
  \left[\nabla\cdot\eta(\mathbf{r})\nabla  - p(\mathbf{r}) \right]G(\mathbf{r},\mathbf{r}')=-4\pi \delta(\mathbf{r}-\mathbf{r}'),
\end{equation}
where $p(\mathbf{r})$ is given by the previous iteration step.  In
three dimensions, we can approximate the equation by,
\begin{equation}
  \mathbf{A}\mathbf{G} = \mathbf{I}, \label{inv}
\end{equation}
where
\begin{equation}
  \mathbf{A}=\frac{1}{4\pi h^3}\left[\mathbf{D} \mathrm{H} \mathbf{D}^T +\mathrm{diag}\{ P \}  \right], \label{green3d}
\end{equation}
$P$ is the vector of function $p(\mathbf{r})$, $\mathbf{G}$ is a
matrix representing lattice Green's function, and $\mathbf{I}$ is unit
matrix.  We see that the solution for the lattice Green's function is
simply equivalent to a matrix inversion,
$\mathbf{G}=\mathbf{A}^{-1}$. Let $C$ be the vector of correlation
function $c(\mathbf{r})$, $\mathbf{G}_0$ be the lattice Green's
function in the free space, and $H_0$ is the vector of dielectric
permittivity defined at integer lattice sites, then we can represent
$C$ by,
\begin{equation}
  C=\mathrm{diag}(\mathbf{G})-\mathrm{diag}(\mathbf{G}_0)/H_0.
\end{equation}
Here $\mathrm{diag}(\cdot)$ is the vector from diagonals of the
argument matrix, to distinguish it from $\mathrm{diag}\{\cdot\}$, and
the division between vectors is Hadamard division (entrywise
division).

\subsection{General remarks on matrix inversion}

The solution of Eq.~\eqref{inv} is the bottleneck of numerical
calculation with the variational field-theoretical equations. One way
of inverting a symmetric and positive definite matrix is by the
Cholesky or LDL factorisation, $\mathbf{A}=\mathbf{L\Lambda L}^T$,
where $\mathbf{L}$ is a lower triangular matrix and $\mathbf{\Lambda}$
is a diagonal matrix; then one computes the inversion
$\mathbf{A}^{-1}$ from $\mathbf{L}^{-1}$ and
$\mathbf{\Lambda}^{-1}$. Since $\mathbf{A}$ is $n^d$-by-$n^d$, the LDL
factorization has complexity $O(n^{3d})$ arithmetic operations and
$O(n^{2d})$ for storage if dense matrix algorithms are used. Both are
prohibitive for multidimensional problems. However, $\mathbf{A}$ is
sparse and has small band width. By optimising the ordering for
evaluating the Cholesky factors, the saving of the computation can be
remarkable \cite{George:SINA:73,DER:SINA:76}. When $d=2$, the
factorisation can be performed with $O(n^3)$ arithmetic operations and
$O(n^2)$ storage. When $d=3$, one requires $O(n^6)$ operations and
$O(n^4)$ storage.

Another property that simplifies the problem is that only the diagonal
of $\mathbf{G}$ is needed for the modified Poisson-Boltzmann
equation. In this case, the complexity for inversion can also be
lowered, because $\mathrm{diag}(\mathbf{A}^{-1})$ can be extracted by
using the Schur complements recursively produced in the intermediate
steps of the LDL factorization.  The recent developed SelInv package
\cite{LYMLYE:ACMMS:11} has provides code to extract the diagonals of a
matrix inverse. It permits the calculation of the diagonals of the
matrix inverse with the same complexity as the factorisation quoted
above.  This gain in efficiency is enough for one to be able to
selectively invert rather large two-dimensional discretisations of
physical problems. The complexity remains rather high for general
three dimensional problems; however, in systems with translational
invariance we will now show how to combine selective inversion with
Fourier analysis.

\subsection{One-dimensional geometry}

Planar interfaces are often studied to investigate the properties
of liquid-liquid interfaces and liquid-solid interfaces of different
systems. We present the dimension reduction for one-dimensional
geometry in this section, which may be useful in, e.g., theoretical
and computational studies of planar electric double layers and
nanopores.

We assume $\eta$ depends only on $z$ coordinate. If the surface
charges are uniformly distributed on planes perpendicular to $x-y$
directions, the potential and the self Green's function will be
translational invariant. Then we can assume $p(\mathbf{r})=p(z)$,
too. The generalized DH equation can be written into,
\begin{equation}
  \nabla\cdot\eta(z)\nabla G(\mathbf{r},\mathbf{r}')=p(z)G(\mathbf{r},\mathbf{r}')-4\pi \delta(\mathbf{r}-\mathbf{r}').
\end{equation}

If we do the polar symmetric Fourier transform to the Green's function
equation in the $x-y$ plane, we obtain the decoupled one-dimensional equation for each
$k$,
\begin{equation}
  \left[\partial_z\eta(z)\partial_z - \eta(z)k^2 -p(z)\right] \hat{G}(k; z,z')= -2 \delta(z-z'), \label{1dg}
\end{equation}
where the coefficient of the delta function becomes $-2$ because the
two-dimensional Fourier transform to $x$ and $y$ comes out a $1/2\pi$.

We know well how to solve Eq.~\eqref{1dg} by discretizing it in a
similar form of \eqref{green3d} such that the one-dimensional lattice
Green's function at spatial frequency $k$ is
\begin{equation}\hat{\mathbf{G}}(k)=2h\left[\mathbf{D} \mathrm{H}
    \mathbf{D}^T +\mathrm{diag}\{ k^2H_0+P \} \right]^{-1},
\end{equation}
with $\mathbf{D}$ and $\mathbf{H}$ being the matrices of one
dimensional case.  On the other hand, we can represent the lattice
Green's function in free space as,
\begin{equation}
  \hat{\mathbf{G}}_0(k)=2h\left[\mathbf{D}\mathbf{D}^T +k^2\mathbf{I} \right]^{-1}. \label{fsgreen}
\end{equation}
The inverse Fourier transform is
$$G(\mathbf{r},\mathbf{r}')=\int_0^\infty \hat{G}(k; z,z')J_0(k\rho)k dk,$$
where $J_0$ is the Bessel function, and thus we find the vector of the correlation function (in the case of $\rho=0$),
\begin{equation}
  C=\int_0^\infty \left[\mathrm{diag}\{\hat{\mathbf{G}}(k)\}-\frac{\mathrm{diag}\{\hat{\mathbf{G}}_0(k)\}}{H_0} \right]  k dk, \label{corr1}
\end{equation}
where the vector division of the second term in the integrand is
Hadamard division.

The free-space Green's function has explicit solution. However, it is
beneficial to calculate it by Eq.~\eqref{fsgreen} in an approximate
way, which gains a singularity cancellation between
$\hat{\mathbf{G}}(k)$ and $\hat{\mathbf{G}}_0(k)$, leading to an
accurate approximation to $C$.

Since the integrand has the property of fast decay, the numerical
integration \eqref{corr1} can be done by a cutoff at a certain frequency $k=K$. We use
a variable transformation $k=e^{\mu v}-1$, where $\mu>0$ is a constant parameter,
and do the integration on $v\in[0, V]$ with $V=\frac{1}{\mu}\log(K+1)$ by a Legendre-Gauss
quadrature, and find a small number of quadrature points can provide high accuracy.

\subsection{Two-dimensional geometry}

Suppose the electric potential and the self Green's function is
uniform in $z$ direction, which is often considered for cylindrical
geometries, e.g., DNA-DNA interaction or charge transport inside
a nanotube. Similar to one dimension, we have
$\eta(\mathbf{r})=\eta(x,y)$ and $p(\mathbf{r})=p(x,y)$, and thus the
modified DH equation becomes,
\begin{equation}
  \nabla\cdot\eta(x,y)\nabla G(\mathbf{r},\mathbf{r}')=p(x,y)G(\mathbf{r},\mathbf{r}')-4\pi \delta(\mathbf{r}-\mathbf{r}'),
\end{equation}
which is essentially still a three-dimensional problem with
coefficients being invariant in $z$ direction.

We can use the Fourier transform in $z$ direction to reduce the
equation into a two dimensional problem. Suppose $\tilde{G}(\omega; x,
y; x', y')$ is the transform of $G$. Then,
\begin{equation}
  \left[\nabla_{xy}\cdot\eta\nabla_{xy} - \eta \omega^2 -p\right] \tilde{G}(k; z,z')= -\sqrt{8\pi} \delta(x-x',y-y'),
\end{equation}
which can be solved by,
\begin{align*}
&\tilde{\mathbf{G}}(\omega)=\sqrt{8\pi}h^2 \mathbf{A}(\omega)^{-1}, ~~~\hbox{and},&\\
&\mathbf{A}(\omega)= \mathbf{D} \mathrm{H} \mathbf{D}^T
+\mathrm{diag}\{ \omega^2H_0+P \},&
\end{align*}
where $\mathbf{D}$ and $\mathbf{H}$ are two-dimensional matrices.
Since the inverse Fourier transform,
\begin{equation}
  G(\mathbf{r},\mathbf{r}')=\frac{1}{\sqrt{2\pi}}\int_{-\infty}^\infty e^{-i\omega z}\tilde{G}(\omega; x, y; x', y')d\omega,
\end{equation}
and the matrix $\mathbf{A}(\omega)$ is symmetric and positive
definite, the lattice Green's function can be represented by the
matrix,
\begin{equation}
  \mathbf{G}=\sqrt{\frac{2}{\pi}}\int_0^\infty \tilde{\mathbf{G}}(\omega) d\omega=4h^2\int_0^\infty \mathbf{A}(\omega)^{-1} d\omega,
\end{equation}
and thus the correlation function is approximated by the integral,
\begin{equation}
  C=\sqrt{\frac{2}{\pi}} \int_0^\infty \left[\mathrm{diag}\{\tilde{\mathbf{G}}(\omega)\}-\frac{\mathrm{diag}\{\tilde{\mathbf{G}}_0(\omega)\}}{H_0} \right] d\omega,
\end{equation}
where $\tilde{\mathbf{G}}_0(\omega)=\sqrt{8\pi}h^2 [\mathbf{D}
\mathbf{D}^T + \omega^2 \mathbf{I}]^{-1} $ is the matrix of lattice
Green's function in free space.

The same method with the Gauss quadrature  as for the one-dimensional case could be adopted
to approximate the integral.

\section{Numerical results}

In this section, we perform numerical study on the convergence and
speed of the proposed algorithm for electrolytes in three-dimensional
space with the presence of one- or two-dimensional charged
interfaces. We set a uniform fugacity parameter $\Lambda=0.2$, and
study the solution with different $\Xi$. In the calculations, the
integral in frequency is discretized with parameters $\mu=1$, $K=32$
and the number of quadrature points between $[0, K]$ is 10, which has been
verified highly accurate for most of examples except for the example
of free energy calculations where 20 quadrature points are used. The correlation function is
subtracted by the bulk value in order to ensure that the bulk property
does not change with the coupling parameter. The error criteria of the
Poisson-Boltzmann solver and the self-consistent iteration are both
$10^{-8}$. Periodic boundary conditions are used for both the PB and the
DH steps. The calculations are performed with a 2.67~GHz Intel 8-core
processor and 48~GB  memory.

\subsection{One planar surface in electrolytes}

We consider one dimensional model in a region $[0, L]$ with $L=32$. In
what follows lengths are measured in the natural scaled units -- the
Gouy-Chapman length, $\ell_{GC}$.  A charged planar interface is
placed at $z=L/2$ separating the electrolyte into two symmetric half
spaces. The surface charge density is $$\rho_f(z)=\delta(z-L/2).$$ The
results of two different coupling parameters $\Xi=1$ and 4 are
calculated. To observe the convergence of the numerical algorithm, the
$n=4096$ solution are taken as the reference solution and the maximum
error $\max|\Phi-\Phi_\mathrm{ref}|$ is
measured. Table~\ref{errorOneplane} displays the errors, the iterative
steps of the self-consistent scheme, and the execution timings of the
two sets of simulations. The ``time1" column represents the execution
times with the selected inversion, the ``time2" is for the direct
inversion of the matrix. The error results show a first-order
convergence of the algorithm since the lattice representation of the
Delta function is first order of accuracy. The number of iterative
steps are significantly increased for a larger $\Xi$, implying more
execution time is needed. For the execution timing, the use of
selected inversion is already remarkable even for one-dimensional
problems, which has several magnitudes of improvement in comparison to
the direct inversion.

The case $n=1024$ corresponds to a rather fine discretisation
$h=\ell_{GC}/32$, which can be expected to resolve the structure of
the electrolyte with high precession.

\begin{table*}[htb]
  \centering
  \caption{Errors, iterative steps, and execution times (in Seconds) for $\Xi=1$ and 4 and $n=128\sim1024$. The case of one planar interface.}
  \begin{tabular}{c | c  c  c  c | c  c  c  c }    \hline
    &    & $\Xi=1$  &    &    &    & $\Xi=4$  &   &     \\ \hline
    $n$   & Error  & steps  & time1  & time2  & Error  & steps  &  time1 & time2   \\ \hline
    128  & 0.115   & 7  & 0.087  &  1.77   & 0.123  & 29  &  0.268  &   6.15 \\ \hline
    256  & 0.058   & 7  & 0.119  &  6.90  & 0.068  & 28  &  0.366  &   25.28 \\ \hline
    512  & 0.027   & 7  & 0.182  & 29.07   & 0.033  & 28  &  0.550  &   104.94  \\ \hline
    1024 & 0.012   & 7  & 0.314  & 121.96  & 0.014  & 27  &  0.921  &   425.36  \\ \hline
  \end{tabular} \label{errorOneplane}
\end{table*}

Next we investigate the free energy of the variational equations with
the same parameters but for a wider range of the coupling
parameter. Appendix \ref{appenb} presents the algorithm for the free
energy calculation. In order to evaluate the determinant of the
lattice Green's function by combining the sparse Cholesky
factorisation with the Fourier analysis, we use the eigenvalues of the
negative Laplace operator at invariant dimensions,\\
$\lambda_{ij}=2\left(1-\cos\frac{\pi
    i}{N+1}\right)+2\left(1-\cos\frac{\pi j}{N+1}\right)$, where
$i,j=1, 2,\cdots, N$ and $N$ is the number of nodes in $x$ and $y$
directions. Suppose $\mb G=\mb A^{-1}$ is the original three
dimensional lattice Green's function. To obtain is determinant which
is the product of all eigenvalues, we decompose the determinant of
$\mb A$ into $N^2$ components of one-dimensional block corresponding
to each given two-dimensional eigenvalue, $(\hat{\mb
  A}+\lambda_{ij})$, where $\hat{\mb A}$ is the discretised matrix in
$z$ direction.  Then the determinant can be evaluated via $\det\mb
A=\Pi_{i,j} \det (\hat{\mb A}+ \lambda_{ij})$.

We use $N=64$ and three meshes, $n=128, 256$ and 512, and a variable
coupling parameter $\Xi=1\sim 4.75$. Table \ref{engOneplane} shows the
results of free energies and iterative steps. It is observed that the
solutions blow up near $\Xi=4.7$ for all three meshes. The finer mesh
only has a slight improvement on the convergence near the blowup
point. The nonphysical blowup phenomenon may be understandable because
the correlation effect between counterions is energetically favorable
for the condensation \cite{Levin:RPP:02}. This implies that, for high
coupling parameter, the variational theory should be further modified;
e.g., by including the excluded-volume effect of the counterions.
  
\begin{table*}[htb]
  \centering
  \caption{Free energies and iterative steps for $\Xi=1\sim 4.75$ and $n=128, 256$ and $512$. The case of one planar interface.}
  \begin{tabular}{c | c  c | c  c | c  c  }    \hline
    &  $n=128$  &   &  $n=256$  &    & $n=512$   &      \\ \hline
    $\Xi$   & Energy  & steps~  & Energy  & steps~ & Energy  & steps~      \\ \hline
    1  & 0.677   & 8   & 0.703  & 8    & 0.715  & 8     \\ \hline
    2  & 0.866   & 12  & 0.891  & 12   & 0.899  & 12    \\ \hline
    3  & 0.944   & 17  & 0.968  & 17   & 0.976  & 17    \\ \hline
    4  & 1.005   & 30  & 1.030  & 30   & 1.037  & 29    \\ \hline
    4.5& 1.049   & 59  & 1.074  & 57   & 1.080  & 52    \\ \hline
    4.6  & 1.065   & 89  & 1.090  & 86   & 1.093  & 70    \\ \hline
    4.65 & 1.079   & 173 & 1.104  & 150  & 1.103  & 92    \\ \hline
    4.70 & blowup  & --  & blowup  & --  & 1.120  & 206   \\ \hline
    4.75 & blowup  & --  & blowup  & --  & blowup  & --   \\ \hline
  \end{tabular} \label{engOneplane}
\end{table*}

\subsection{Two planar interfaces with a sandwiched layer}

The second example is the case of two charged planar interfaces
separating the space into three parts. The part between the interfaces
is low dielectric region, and the two side parts are
electrolytes. The computational interval is $[0, L]$ with $L=32$,
where the region of $[0.4L, 0.6L]$ is inaccessible to ions. The
dielectric ratio is set to be $\eta(z)=0.1$ in the region $[0.42L,
0.58L]$ and 1 otherwise, and we see there is a $0.02L$ thickness of
buffer zone between the ionic fluid and the dielectric object. This
dielectric medium is a model of membrane.  We suppose asymmetric
unit surface charges are placed at the two ends of the ion
inaccessible region, and thus the fixed charge density is
$$\rho_f(z)=\delta(z-0.4L)-\delta(z-0.6L).$$

As in the previous example, we first calculate the errors and
execution timings with the mesh refinement for two coupling parameters
$\Xi=1$ and 4, shown in Table~\ref{errorTwoplane}. The first order of
accuracy remains, and the convergence of self-consistent iteration is
rapid for small $\Xi$, but becomes worse with the increase of the
coupling parameter. Mesh refinement slightly improves the convergence
rate.

\begin{table}[h]
\centering
  \caption{Errors, iterative steps, and execution times (in Seconds) for $\Xi=1$ and 4 and $n=128\sim1024$. The case of two planar interfaces. }
  \begin{tabular}{c | c  c  c   | c  c  c  }    \hline
    &    & $\Xi=1$  &    &    & $\Xi=4$  &        \\ \hline
    $n$   & Error  & steps  & time1   & Error  & steps  &  time1   \\ \hline
    128  & 0.285   & 8  & 0.184  & 0.344  & 83  &  0.861        \\ \hline
    256  & 0.139   & 8  & 0.265  & 0.145  & 45  &  0.712      \\ \hline
    512  & 0.082   & 8  & 0.435  & 0.024  & 40  &  1.031      \\ \hline
    1024 & 0.028   & 7  & 0.753  & 0.012  & 36  &  1.725      \\ \hline
  \end{tabular} \label{errorTwoplane}
\end{table}

\begin{figure}[htbp]
\centering
  \includegraphics[width=0.45\textwidth]{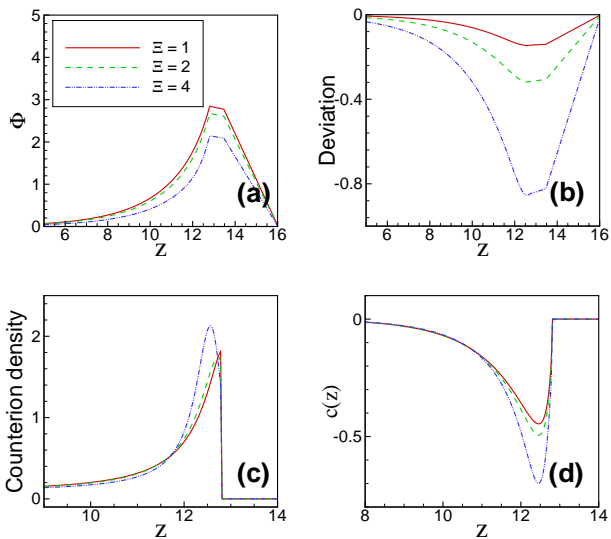}
  \caption{Results of $\Xi=1, 2,$ and 4 for the case of two planar
    interfaces. (a) Potential distributions; (b) Potential deviations
    from the Poisson-Boltzmann solutions; (c) Counterion charge
    density; (d) Correlation functions. } \label{DMX}
\end{figure}

We then fix the number of grid points, $N=1024$, and calculate the
results of $\Xi=1, 2$ and 4. In Fig.~\ref{DMX}, we plot the potential
distributions, their deviations from the Poisson-Boltzmann solution,
the counterion charge densities, and the correlation functions for
these three $\Xi$. Only the left half solutions are plotted due to the
symmetry or antisymmetry along $z=L/2$. From (a-b), it is observed that with the
increase of coupling parameter, the potential magnitude becomes
small. It is reasonable in the sense the correlation function
$c(\mathbf{r})$ is negative (Fig.~\ref{DMX}(d)) and favorable for high
counterion concentration. At $\Xi=2$, we see the deviation of the
self-consistent equations from the Poisson-Boltzmann equation is
already large, which is over $10\%$ difference at the maximum
potential.  Another phenomenon we can observe from (c) is that, with
the increase of $\Xi$, the image charge effect is strengthened, which
behaviors a repulsive interaction to the counterions.

\subsection{Circular geometry}

We consider an example of 2D geometry of $L^2$ with $L=32$ in units of
the Gouy-Chapman length.  The electrolyte fills the space outside a
circle of radius $R$. Unit line charge density is placed on this
circle, but it has distinct signs in each half surface, i.e.,
$$\rho_f(x,y)=\mathrm{sign}(y)\delta(\sqrt{x^2+y^2}-R).$$
This setting models a Janus particle.  The dielectric ratio $\eta$
equals $0.1$ inside a circle of radius $R-0.1$, and 1 outside. This
$0.1$ gap characterizes the size of mobile ions, and eliminates the
singularity of computing the correlation function in DH steps.

\begin{figure} [htbp]
\centering
  \includegraphics[width=0.25\textwidth]{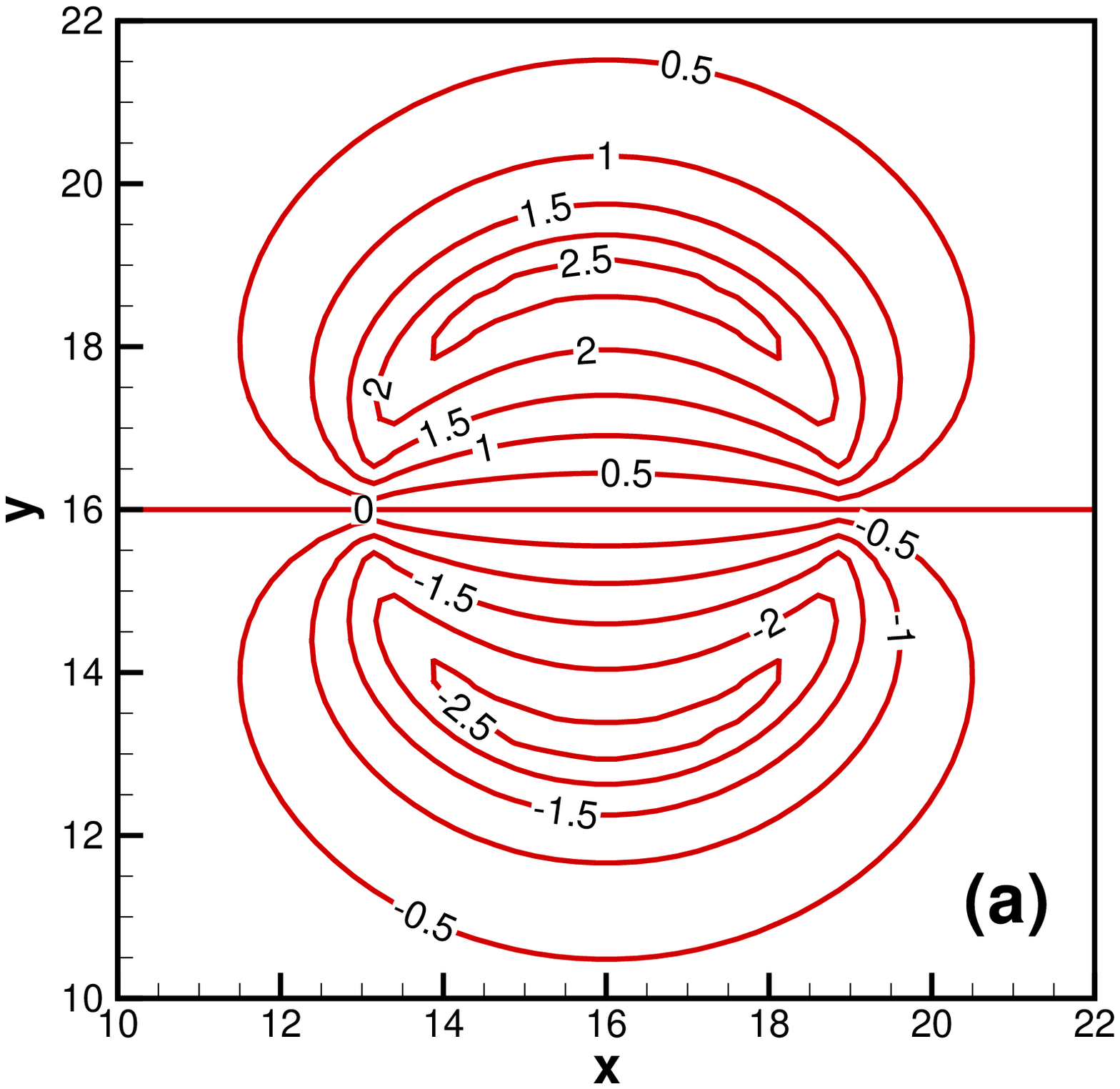}\includegraphics[width=0.25\textwidth]{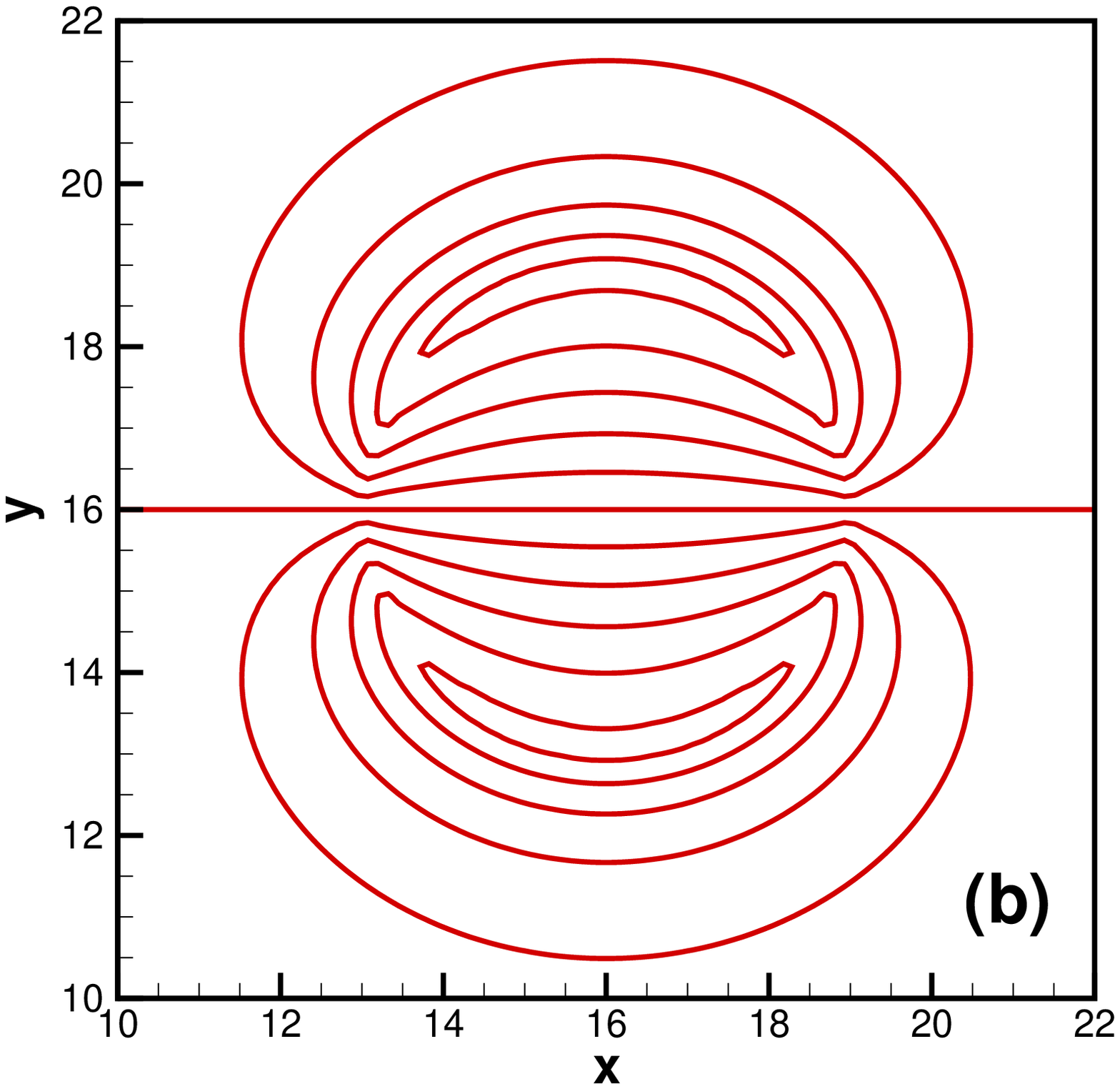}\\
  \includegraphics[width=0.25\textwidth]{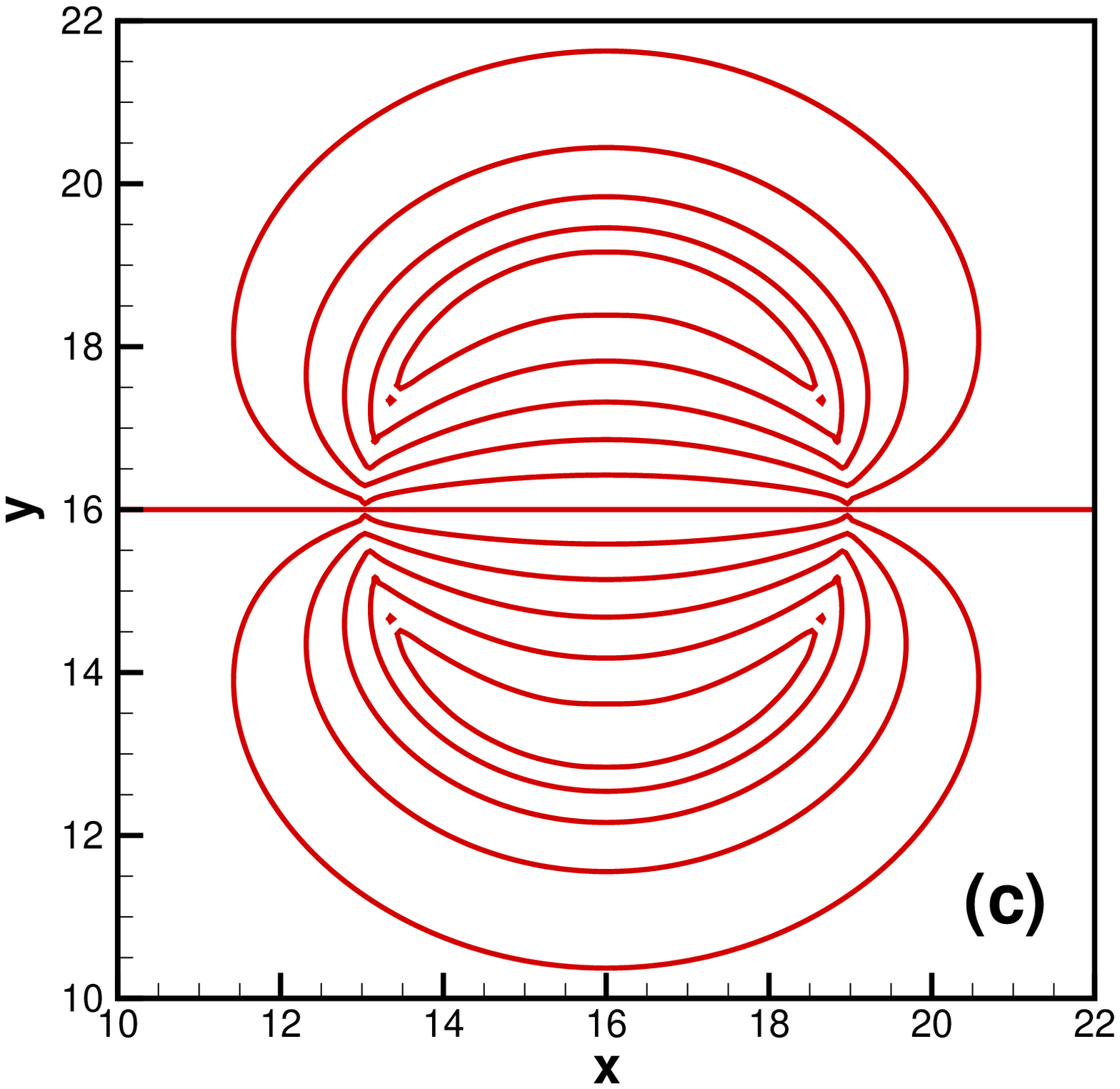}\includegraphics[width=0.25\textwidth]{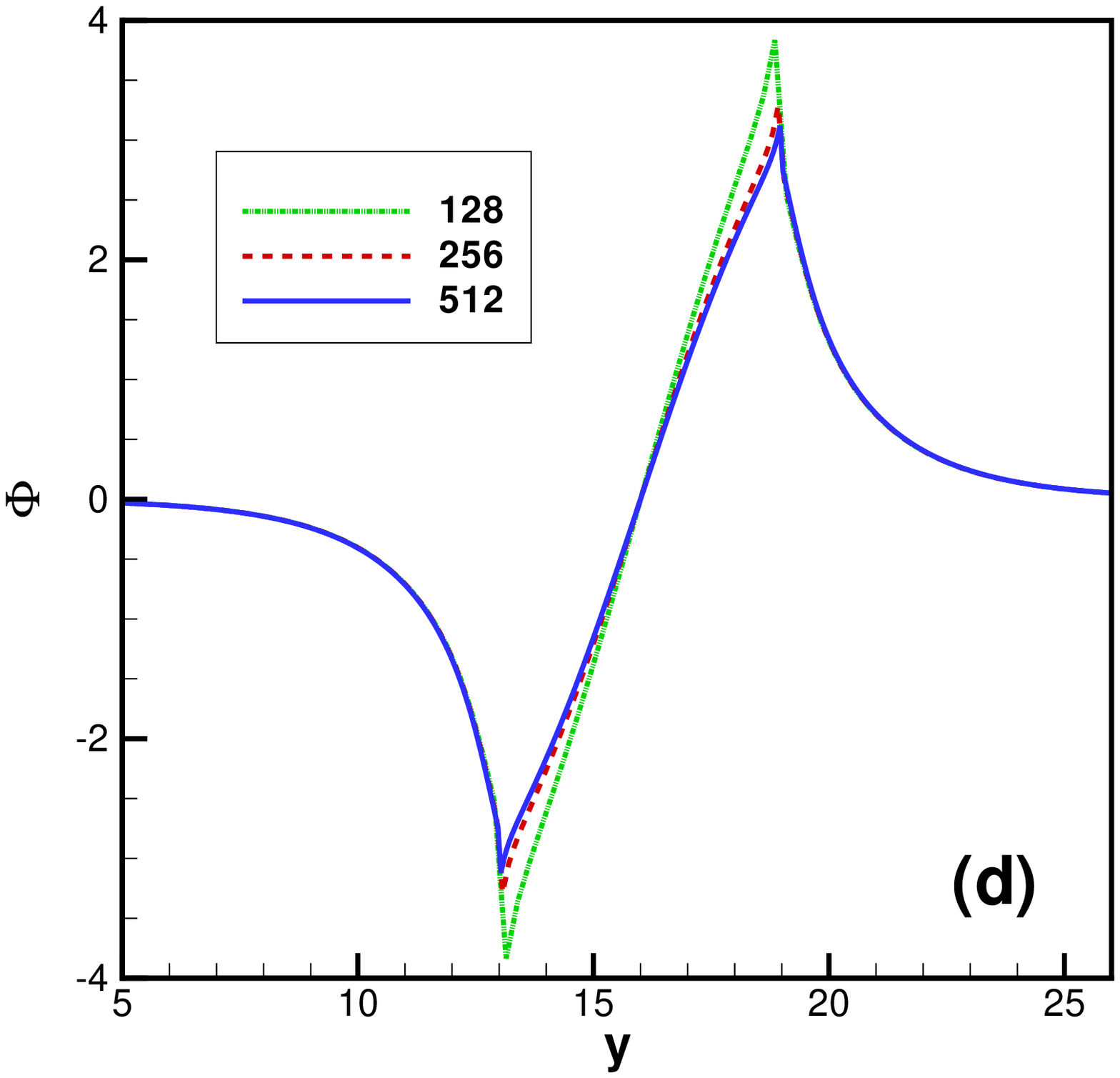}
  \caption{Potential contours of the two dimensional case for (a) $128^2$
    mesh, (b) $256^2$ mesh, and (c) $512^2$ mesh; and (d) the potentials at the $x=0$ line for the three meshes.} \label{twod}
\end{figure}

To see the convergence and execution timings of the algorithms for two
dimesions, we take the coupling parameter $\Xi=4$, and three mesh sizes,
$128^2$, $256^2$ and $512^2$.  The surface charge is approximated by 128
cations and 128 anions, which are then distributed onto lattice
sites. Fig.~\ref{twod} plots the contours of the potential
distributions of three cases, and their potential values at the $x=0$ line,
which show the convergence of the algorithm. Table~\ref{t2d} lists the potential
difference between the current step and the previous step at each
iteration, and the execution times of the PB step and the DH
step. Only the data of the first 9 iterative steps are given, and
overall they are convergent both after $\sim30$ steps. It is found that the
PB sub-step is very fast after a few iterative steps and most of
execution time is spent on the DH steps. For $256^2$ and $512^2$ meshes, each
iterative step for the single processor running can be accomplished in
$\sim 1$ minute and $\sim 11$ minutes, respectively, even although dozens of selected inversion
operations must be performed in order to perform the Fourier integral.

\begin{table*}[htbp]
  \caption{Potential differences and execution times (in Seconds) of the PB and DH steps in the first 9 iterative steps. $128^2$, $256^2$ and $512^2$ meshes. }
  \begin{tabular}{c | ccc | ccc  }    \hline
     && $128^2$ mesh      &          & & $256^2$ mesh   &         \\ \hline
    ~Step~ & Diff & PB step & DH step& Diff & PB step & DH step   \\ \hline
    0 &  3.88336 &   8.29 &   3.92   & 3.32273 & 45.09  & 56.48   \\ \hline
    1 &  0.15133 &   1.22 &   4.22   & 0.11065 &  6.25  & 56.54  \\ \hline
    2 &  0.03013 &   1.19 &   4.08   & 0.03376 &  5.85  & 56.54  \\ \hline
    3 &  0.01704 &   1.06 &   4.18   & 0.01860 &  5.31  & 56.39  \\ \hline
    4 &  0.00888 &   1.08 &   3.92   & 0.00966 &  5.37  & 56.43  \\ \hline
    5 &  0.00454 &   1.02 &   4.26   & 0.00498 &  4.93  & 58.24  \\ \hline
    6 &  0.00235 &   1.00 &   4.29   & 0.00260 &  5.03  & 58.34  \\ \hline
    7 &  0.00123 &   0.90 &   3.92   & 0.00138 &  4.53  & 58.24  \\ \hline
    8 &  0.00065 &   0.89 &   3.92   & 0.00074 &  4.53  & 56.99  \\ \hline
  \end{tabular}
   \begin{tabular}{c | ccc }    \hline
      &        & $512^2$ mesh &          \\ \hline
  ~Step~ &  Diff & PB step   & DH step      \\ \hline
    0 &  3.10677 & 239.88  & 666.14      \\ \hline
    1 &  0.11336 &  34.88  & 669.50     \\ \hline
    2 &  0.03271 &  32.97  & 669.89     \\ \hline
    3 &  0.01787 &  31.42  & 672.02     \\ \hline
    4 &  0.00916 &  30.77  & 669.03     \\ \hline
    5 &  0.00466 &  29.16  & 667.80     \\ \hline
    6 &  0.00240 &  28.16  & 672.96     \\ \hline
    7 &  0.00126 &  27.41  & 669.31     \\ \hline
    8 &  0.00067 &  24.82  & 668.74     \\ \hline
  \end{tabular} \label{t2d}
\end{table*}

We note that we have been unable to use our algorithm to study systems
roughly with $\Xi >5$. We have already shown that convergence is {\it slower}
when $\Xi$ increases, however at some point we find that the iterative
process does not converge. This seems to be associated with a
condensation of charges to the surface. It is possible that including
finite size effects for the ions could prevent this phenomenon, but we
leave further study of this point to a later study.

\section{Conclusion}

We have developed an algorithm for solving fluctuation enhanced
Poisson-Boltzmann equations derived using a variational field-theoretic
approach. By studying numerical examples, we show the algorithm has
attractive performance for both accuracy and efficiency.

For the moment considerations of speed limit detailed studies to one-
and two-dimensional geometries.  Extension of the algorithm to large
three-dimensional geometries is straightforward, but limits to a small mesh
(our single-processor machine allows the computation with the $64^3$ mesh
in a few hours). Exploring the finer mesh solution requires a parallel
implementation of the selected inversion algorithm.

In the framework of self-consistent equations, we aim to include
the Born energy contribution to the self energy for
charged systems in smoothly varying dielectric media
\cite{Wang:PRE:10} in order to study more complex, heterogeneous
materials.

\appendix

\section{Variational derivation in matrix form}\label{appena}

In this appendix we give a derivation of the variational equations
starting from a discretised version of the free energy. This
derivation is useful because it removes many of the ambiguities and
divergences present in the continuum formulation.

Suppose $\Phi$ and $\rho_f$ are vectors, $\mathbf{G}$ and
$\mathbf{G}_0$ are matrices, and $-\mathbf{D}^T$ is the discrete
correspondence of gradient operator. The the matrix operator $\mathbf
D$ is discretised version of the divergence. Functions of vectors,
such as $\cosh\Phi$, correspond to a vector made of the individual
elements, in a manner familiar in Matlab. Then in discrete form, the
field-theoretic Gibbs free energy can be written as,
\begin{align}
\mathcal{F}_\mathrm{Gibbs}&=
-\frac{1}{2}\mathrm{Tr} \log (\Xi\mathbf{G})-
 \frac{ (\mathbf{D}^T\Phi)^T {\bf H} \mathbf{D}^T\Phi}{8\pi\Xi}  \nonumber \\
&+ \mathrm{Tr~diag} \left\{ \frac{\rho_f\circ \Phi}{2\pi\Xi} \right\} -
 \mathrm{Tr}\frac{\mathbf{D} \mathbf{H D}^T\mathbf{G}}{8\pi} \nonumber \\
&-\frac{\Lambda}{4\pi\Xi}\mathrm{Tr~diag} \left\{  e^{- \Xi \Delta G_{rr}/2}\circ\cosh\Phi \right\},
\end{align}
where ``$\circ$" is the Hadamard product, and  $\Delta
G_{rr}=\mathrm{diag}(\mathbf{G}-\mathbf{G}_0)$ represent the
difference vectors of diagonal elements of $\mathbf{G}$ and
$\mathbf{G}_0$. We use matrix derivatives with respect to vector or
matrix (e.g., refer to \cite{PP:book:08}) for the variation with
respect to $\Phi$ and $G$.

The variation with respect to  $\Phi$ gives,
\begin{align}
&\frac{\delta \mathcal{F}_\mathrm{Gibbs}}{\delta \Phi}=-\mathbf{D} \frac{\partial \mathcal{F}_\mathrm{Gibbs}}{\partial \mathbf{D}^T \Phi} + \frac{\partial \mathcal{F}_\mathrm{Gibbs}}{\partial \Phi}  \nonumber \\
&~~=-\frac{\mathbf{DHD}^T\Phi}{4\pi\Xi} + \mathrm{diag} \left\{ \frac{\rho_f}{2\pi\Xi} -\frac{\Lambda}{4\pi\Xi}  e^{- \Xi \Delta G_{rr}/2}\circ\sinh\Phi \right\}. \nonumber \\
\end{align}
Using $4\pi\Xi\delta \mathcal{F}_\mathrm{Gibbs}/\delta \Phi=0$ yields the modified Poisson-Boltzmann equation,
\begin{equation}
-\mathbf{DHD}^T\Phi + \mathrm{diag} \left\{ 2 \rho_f  - \Lambda  e^{- \Xi \Delta G_{rr}/2}\circ\sinh\Phi \right\}=\mathbf{0}.
\end{equation}

Similarly, the variation of the free energy to matrix $\mathbf{G}$ is,
\begin{eqnarray}
&&\frac{\delta \mathcal{F}_\mathrm{Gibbs}}{\delta \mathbf{G}} =- \mathbf{D} \frac{\partial \mathcal{F}_\mathrm{Gibbs}}{\partial \mathbf{D}^T \mathbf{G}}  + \frac{\partial \mathcal{F}_\mathrm{Gibbs}}{\partial \mathbf{G}}  \nonumber\\
&&=  \mathbf{D} \frac{\mathbf{H D}^T  \mathbf{I}}{8\pi}  -  \frac{1}{2\mathbf{G}} +\frac{\Lambda}{8\pi}\mathrm{diag} \left\{  e^{- \Xi \Delta G_{rr}/2}\circ\cosh\Phi \right\}. \nonumber \\
\end{eqnarray}
By  $\delta \mathcal{F}_\mathrm{Gibbs}/\delta \mathbf{G} =\mathbf{0}$ and right multiplying by $8\pi \mathbf{G}$, we find the generalized Debye-H\"uckel equation,
\begin{equation}
\mathbf{DHD}^T \mathbf{G }+  \mathrm{diag} \left\{\Lambda   e^{- \Xi \Delta G_{rr}/2}\circ\cosh\Phi \right\}\mathbf{G}= 4\pi\mathbf{I}.
\end{equation}

\section{Evaluation of the self-consistent free energy, divergences} \label{appenb}

The continuum limit of the energy eq.~\eqref {gibbs} in three
dimensions is divergent in $ 1/h^3$ as the cut-off in the problem, $h$,
is taken to zero. In this appendix we show that this problem is easily
solved by always calculating energies compared to an empty system with
$\rho_f=0$ and $\Lambda=0$. There is also another small problem with
the use of eq.~\eqref{gibbs} in a working code -- it requires access
to values of $G$ which are near to, but not on the diagonal in order
to calculate the derivatives. Selected inversion does not easily give
this information so we eliminate the derivative using the variational
equations eq.~\eqref{dimensionless}:
\begin{align*}
\mathcal{F}= -& \frac{1}{2} {\rm Tr} \log(\Xi G) - \int d\r\left [ \frac{\eta (\nabla\Phi)^2 }{8 \pi \Xi} - \frac{\rho_f
  \Phi}{2 \pi \Xi}  \nonumber
 \right ] \\ -& \int d\r \frac{\Lambda \chi}{8 \pi \Xi}  \expG
\cosh \Phi ~ G(\r,\r) \nonumber
\\ -& \int d\r \frac{\Lambda \chi}{4 \pi \Xi}
\expG \cosh \Phi
\end{align*}
where we have dropped a constant coming from the integral of a
$\delta$-function. We now shift the zero of free energy taking as a
reference,
\begin{equation}
\mathcal{F}_\mathrm{ref} = \mathcal{F}(\rho_f=0, \Lambda=0),
\end{equation}
for the system\begin{equation}
-\nabla \cdot \eta \nabla G_\mathrm{ref} = 4 \pi  \delta( \r -\r' )
\end{equation}
If $\eta=1$ everywhere then $G_\mathrm{ref}=G_0$; in a more general background the
functions are different.

Firstly we study the behavior of $\mathcal{F}-\mathcal{F}_\mathrm{ref}$. Let us consider only the
potentially divergent terms:
\[
  -\frac{1}{2} \Tr \log(G/G_\mathrm{ref})
- \frac{\Lambda \chi }{8  \pi \Xi} \expG \cosh\Phi~ G(\r ,\r ).
\]
The last term containing $G(\r,\r)$ has a naive divergence in $1/h^3$. We
now show that this divergence is cancelled by a residual divergence in
the logarithmic contribution:
\begin{align*}
-\frac{1}{2}\log(G/G_\mathrm{ref}) =& \frac{1}{2}\log \left( \frac{-\nabla^2 + \Lambda \chi \cosh \Phi
    \expG }{-\nabla^2 } \right )\\
 =& \frac{1}{2}\log \left( 1 + \frac{G_\mathrm{ref}} {4 \pi \Xi} \Lambda \chi \cosh \Phi  \expG
 \right ). \nonumber
\end{align*}
Take the trace of this expression. When the eigenvalues of the
expression involving $G_\mathrm{ref}$ are small we can expand the log in a Taylor
series to find a contribution:
\begin{equation}
\frac{G_\mathrm{ref}} {8 \pi \Xi} \Lambda \chi \cosh\Phi ~ \expG
\end{equation}
We find that the dangerous parts combine to give:
\begin{equation}
\frac{\Lambda \chi }{8 \pi \Xi} \expG \cosh\Phi  \left[ G (\r,\r) -
  G_\mathrm{ref}(\r,\r) \right].
\end{equation}
The combination $G-G_\mathrm{ref}$ should remain finite as the mesh spacing goes
to zero. Thus $\mathcal{F}-\mathcal{F}_\mathrm{ref}$ does too.

Using the variational identity eq.~\eqref{dimensionless} we now find:
\begin{align*}
&\mathcal{F}-\mathcal{F}_\mathrm{ref} = - \frac{1}{2} {\rm Tr} \log(G/G_\mathrm{ref}) +
\frac{1}{4 \pi \Xi }\int d\r
  \rho_f  \Phi + \\
&
\int d\r
 \frac{\Lambda \chi \expG} {8 \pi \Xi}
\left [
 \Phi \sinh{\Phi}
- \cosh{\Phi}\, ~ \Xi G(\r,\r)
-2  \cosh{\Phi}
\right ].
\end{align*}
This is our required result: A free energy which remains finite as the
cut-off is reduced, and which does not require access to non-diagonal
elements of $G$. Since ${\rm Tr} \log =\log \det$, the determinants are calculated rather easily through
Cholesky factorisation of the corresponding sparse operator \cite{pasquali:2008}.

\section*{Acknowledgement}

Z. Xu acknowledges the financial support from the Natural Science
Foundation of China (Grant Numbers: 11101276 and 91130012) and the
Alexander von Humboldt foundation. A.C. Maggs is supported in part by
the agence nationale de la recherche via the project FSCF.

\bibliographystyle{elsart-num} 
\bibliography{biobib}

\end{document}